\documentclass[aps, showpacs, showkeys,nofootinbib,floatfix]{revtex4}

\usepackage{amssymb}
\usepackage{amsmath}
\usepackage{graphicx}
\usepackage{hyperref}

\begin{document}

\title{Constraints on accelerating universe using ESSENCE and Gold supernovae data  combined with other cosmological probes }

\author{Jianbo Lu}
\email{lvjianbo819@163.com}
\author{Lixin Xu}
\author{Molin Liu}
\author{Yuanxing Gui}

\affiliation{School of Physics and Optoelectronic Technology, Dalian
University of Technology, Dalian, 116024, P. R. China}

\begin{abstract}

We use recently observed data:  the 192  ESSENCE type Ia supernovae
(SNe Ia), the  182 Gold SNe Ia, the 3-year WMAP,
 the SDSS baryon acoustic peak, the X-ray gas
mass fraction in clusters  and the observational $H(z)$ data to
 constrain models of the accelerating universe.
 Combining the 192 ESSENCE data with the observational $H(z)$ data to
constrain a parameterized deceleration parameter, we obtain the best
fit values of transition redshift and current deceleration parameter
$z_{T}=0.632^{+0.256}_{-0.127}$, $q_{0}=-0.788^{+0.182}_{-0.182}$.
Furthermore, using $\Lambda$CDM model and two model-independent
equation of state of dark energy, we find that the combined
constraint from the 192 ESSENCE data and other four cosmological
observations gives smaller values of $\Omega_{0m}$ and $q_{0}$, but
a larger value of $z_{T}$ than the combined constraint from the 182
Gold data with other four observations. Finally, according to the
Akaike information criterion it is shown that the recently observed
data equally supports three dark energy models: $\Lambda$CDM,
$w_{de}(z)=w_{0}$ and $w_{de}(z)=w_{0}+w_{1}\ln(1+z)$.

\end{abstract}


\maketitle

\section{$\text{Introduction}$}

 {\small {~~~}}~The type Ia supernovae (SNe Ia)
investigations \cite{1Riess}, the cosmic microwave background(CMB)
results from WMAP \cite{2Spergel} observations, and  surveys of
galaxies \cite{3Pope} all suggest that the expansion of present
universe is speeding up rather than slowing down. If one considers
that the evolution of universe complys with the standard cosmology,
the accelerated expansion of the present universe is usually
attributed to the fact that dark energy (DE) is an exotic component
with negative pressure. Many kinds of DE models have already been
constructed such as $\Lambda$CDM \cite{4Weinberg}, quintessence
\cite{5Ratra}, phantom \cite{6Caldwell}\cite{07Setarea}, generalized
Chaplygin gas (GCG) \cite{7Kamenshchik}\cite{7zhu}, modified
Chaplygin gas \cite{8Benaoum}\cite{9Lu}\cite{9LuMCG2}, quintom
\cite{10Feng}, holographic dark energy \cite{11Li}\cite{12Zhang},
agegraphic dark energy\cite{07Cai}\cite{08Zhang},  and so forth.

On the other hand, to remove the dependence of special properties of
extra energy components, a parameterized  equation of state (EOS) is
assumed for DE. This is also commonly called the model-independent
method. The parameterized EOS of dark energy $w_{de}$ which is
popularly used in parameter best fit estimations, describes the
possible evolution of DE. For example, $w_{de}(z)=w_{0}$= const
\cite{13Hannestad}, $w_{de}(z)=w_{0}+w_{1} \ln(1+z)$ \cite{14Gerke}.
 The
parameters $w_{0}$, $w_{1}$ are obtained by the best fit estimations
from cosmic observational datasets.

Recently, the 192 ESSENCE SNe Ia data \cite{15Wu} was compiled by
Ref. \cite{16Davis} using the four sets of supernova (SN) data:  60
ESSENCE SNe \cite{21Wood-Vasey},  57 Supernova Legacy Survey (SNLS)
SNe \cite{39Astier},  45 nearby SNe
\cite{1Riess}\cite{40Hamuy}\cite{41Jha},  and 30 new SNe at high
redshift ($0.216 \leq z \leq 1.755$) recently discovered by the
Hubble Space Telescope (HST) and classified as $^{"}$Gold$^{"}$ SNe
by Ref. \cite{17Riess}. The ESSENCE project \cite{21Wood-Vasey} is a
ground-based survey that design to detect about 200 SNe Ia in the
range of $z = 0.2-0.8$ to measure the EOS of DE to better than 10
percent.
 The SNLS  and the nearby
SNe data as the complementary cosmological probes have been refitted
by \cite{21Wood-Vasey} with the same lightcurve fitter used for the
ESSENCE data. As regards the 30 HST SNe, it is necessary to perform
a normalization\cite{07042606Lazkoz}.  Ref. \cite{16Davis} adopted
the low redshift SNe that these samples had in common in order to
normalize the luminosity distances of the samples, and the error in
the normalization is included in the distance errors for the HST SNe
\cite{16Davis}\cite{07042606Lazkoz}.

In Ref. \cite{16Davis} the authors applied the 192 ESSENCE SNe Ia
data, the 3-year WMAP CMB shift parameter
\cite{26Spergel}\cite{18Wang}, the baryon acoustic oscillation (BAO)
peak from Sloan Digital Sky Surver (SDSS) \cite{19Eisenstein} to
constrain the current values of EOS of DE $w_{0de}$ and
dimensionless matter density $\Omega_{0m}$ by using several
model-independent EOS of DE. However, some other cosmological
quantities  such as transition redshift $z_{T}$ and current
deceleration parameter $q_{0}$ were not discussed. On the other
hand, we know that the 182 Gold SNe Ia data \cite{17Riess} is
compiled from five distinct subsets defined by the group or
instrument that discovered and analyzed the corresponding SNe data.
These subsets are \cite{28Nesseris}\cite{30website182subset}: the
High $z$ Supernova Search Team (HZSST) subset (41 SNe)
\cite{1Riess}\cite{HZSST2}\cite{HZSST3}\cite{HZSST4}, the Supernova
Cosmology Project (SCP) subset (26 SNe) \cite{SCP}, the Low Redshift
(LR) subset (38 SNe) \cite{40Hamuy}\cite{LR2}\cite{28Jha}\cite{LR4},
the HST subset (30 SNe) \cite{17Riess} and the SNLS subset (47 SNe)
\cite{39Astier}. It can be found that there are 99 SNe that are in
the 192 ESSENCE data but not
 in the 182 Gold data\footnote{The two sets of SNe Ia data with their
 subsets
 are shown in the
Appendix. From the Appendix it can be seen that there are 93 SNe Ia
in common between the 192 ESSENCE data and the 182 Gold data(i.e.,
from number 81 to number 173 in  TABLE IV). They include 25 nearby
SNe (or 25 LR SNe), 30 HST SNe and 38 SNLS SNe.}. Furthermore,
relative to the  Gold sample  Ref. \cite{16Davis}\cite{21Wood-Vasey}
applied
 an updated version of the MLCS2k2
  method\footnote{The basic framework for Multicolor
Light Curve Shape method to measure the luminosity distances was
laid out
 by Ref. \cite{28Jha}  in 2002 (i.e., MLCS2k2 method) and it has already been applied
to SN Ia cosmology such as the 157 Gold  SNe Ia data \cite{22Riess}
and the 182 Gold  SNe Ia data  \cite{17Riess}. A  new version of the
MLCS2k2 was developed with an expanded training set by Ref.
\cite{41Jha} and this light-curve fitting technique has also been
applied to measure the luminosity distances to  ESSENCE, SNLS and
nearby SNe Ia in Ref. \cite{21Wood-Vasey}. Because the basic MLCS2k2
algorithms were designed  by Ref. \cite{28Jha}, Refs.
\cite{21Wood-Vasey}\cite{41Jha} continue to refer to this updated SN
distance fitter as MLCS2k2, even though its implementation,
applicability, and robustness have evolved substantially since then.
For more details about MLCS2k2 please see Refs.
\cite{21Wood-Vasey}\cite{41Jha}\cite{17Riess}\cite{28Jha}\cite{22Riess}\cite{MLCS2k2website}.}
to measure distances to SNe Ia for the  ESSENCE sample,
incorporating new procedures for $K$-correction and extinction
corrections. So,
 the data points are also different even though  for the
same SNe in the two SN  samples\footnote{For the case of the 192
ESSENCE data, since the error in the normalization is included in
the distance errors for the HST SNe,  the data points from the   30
HST SNe are also different  between the 192 ESSENCE data and the 182
Gold data.}. Therefore, we want to know what are the differences for
the constraints on cosmological quantities from these two samples of
SNe Ia respectively. In this paper, by using a parameterized
deceleration parameter and model-independent EOS of DE, we apply the
recent cosmic observations to constrain several cosmological
quantities, such as $z_{T}$, $q_{0}$, and compare the differences
for them when the constraints are obtained from  the 192 ESSENCE
data and  the 182 Gold data, respectively. To avoid the degeneration
of DE models and get the significant constraints on cosmological
quantities, we combine other observational data with these two sets
of SNe data, such as the 3-year WMAP CMB shift parameter, the BAO
peak from SDSS, the X-ray gas mass fraction in clusters
\cite{20Allen} and the observational $H(z)$ data from the Gemini
Deep Deep Survey (GDDS) \cite{23Abraham} and archival data
\cite{24Treu}\cite{25Treu}.

 The paper is organized as follows. In section 2, we apply recent cosmic observations to constrain models of the
 accelerating
 universe by using a parameterized deceleration parameter $q$ and model-independent EOS of
 DE
$w_{de}$. In section 3, we use the information criterion of model
selection for DE models to estimate which model for an accelerating
universe is distinguish by statistical analysis of observational
datasets. Section 4 is the conclusion.

\section{$\text{Constraining models of the accelerating universe}$}
\label{section2}
\begin{normalsize}
\begin{center}
\bfseries 2.1 Constraining models of the accelerating universe
using a parameterized deceleration parameter
\end{center}
\end{normalsize}

The advantage of parameterizing $q(z)$ is that the conclusion does
not depend on any particular gravitational theory \cite{35Gong}. In
this section we consider a parameterized deceleration parameter
$q(z)=\frac{1}{2}+\frac{a+bz}{(1+z)^{2}}$
\cite{35Gong}\cite{15Gong}, where $a$, $b$ are constants. This
deceleration parameter may have the same behavior as the simple
three-epoch model \cite{35Gong}. Originally, the 157 Gold SNe Ia
data was applied to constrain the transition redshift $z_{T}$ by
parameterizing
 a deceleration parameter $q(z) = q_{0} + q_{1}z$ in Ref.
 \cite{22Riess},
and the result was given as  $z_{T}=0.46\pm0.13$
($1\sigma)$\footnote{In this  paper, all errors are $1\sigma$
statistical errors.}. However, it was soon realized that such a
parametrization can not re-produce the behavior of the cosmological
constant \cite{14Virey}. An alternative parametrization is  a simple
three-epoch model of $q(z)$ \cite{11Turner}\cite{15Shapiro}, where
the function $q(z)$ is not smooth. Since the current SN Ia data is
still sparse, the division of the data to three different redshift
bins may not be a good representation of the data \cite{35Gong}.
Then following Ref. \cite{15Shapiro}, the authors in Ref.
\cite{35Gong} proposed a simple smooth function
$q(z)=\frac{1}{2}+\frac{a+bz}{(1+z)^{2}}$ which is more realistic
and then used the 157 Gold  SNe Ia data  to constrain the  $z_{T}$.
In this paper using this parameterized deceleration parameter, we
also want to know what are the best fit values of $z_{T}$ and
$q_{0}$ from the latest 192 ESSENCE SNe Ia data, and what are the
differences for the constraints on $z_{T}$ and $q_{0}$ when compare
them  with the constraint from Gold SNe Ia data. Next we will
discuss these questions.

According to the definition of the Hubble parameter
$H(t)=\frac{\stackrel{.}{a}}{a}$ and the deceleration parameter
$q(t)=-\frac{\stackrel{..}{a}}{aH^{^{2}}}$, we get
\begin{equation}
H(z)=H_{0} \exp[\int^{z}_{0}[1+q(u)]d \ln(1+u)].\label{4H}
\end{equation}
Substituting the expression
$q(z)=\frac{1}{2}+\frac{a+bz}{(1+z)^{2}}$ into Eq. (\ref{4H}), we
obtain
\begin{equation}
H^{2}(z)=H_{0}^{2}E^{2}(z)=H_{0}^{2}(1+z)^{3}
\exp[\frac{2az+(a+b)z^{2}}{(1+z)^{2}}].\label{5H2}
\end{equation}

Since type Ia Supernovae behave as Excellent Standard Candles, they
can be used to directly measure the expansion rate of the universe
up to high redshift  for comparison with the present rate.
Therefore, they provide direct information on the universe$^{,}$s
acceleration and constrain the DE model. Theoretical dark energy
model parameters are determined by minimizing the quantity
\begin{equation}
\chi^{2}_{SNe}(H_{0}, \theta)=\sum_{i=1}^{N}\frac{(\mu_{obs}(z_{i})
-\mu_{th}(H_{0},\theta,z_{i}))^2}{\sigma^2_{obs;i}},\label{1chi2-sne}
\end{equation}
where $N=192$ for the ESSENCE SNe Ia data \cite{16Davis},
$\sigma^2_{obs;i}$ are errors due to flux uncertainties, intrinsic
dispersion of SNe Ia absolute magnitude and peculiar velocity
dispersion respectively. $\theta$ denotes model parameters. The
theoretical distance modulus $\mu_{th}$ is defined as
\begin{equation}
\mu_{th}(z_{i})\equiv
m_{th}(z_{i})-M=5log_{10}(D_{L}(z))+5log_{10}(\frac{H_{0}^{-1}}{Mpc})+25,\label{2mu-th}
\end{equation}
where
\begin{equation}
D_{L}(z)=H_{0}d_{L}(z)=(1+z)\int_{0}^{z}\frac{H_{0}dz^{'}}{H(z^{'};\theta)},\label{3dl}
\end{equation}
 $\mu_{obs}$ is given by supernovae dataset, and $d_{L}$ is the
luminosity distance. $H$ is the Hubble parameter, $^{"}$0$^{"}$
denotes the current value of the variable.

\begin{figure}[!htbp]
  \includegraphics[width=6cm]{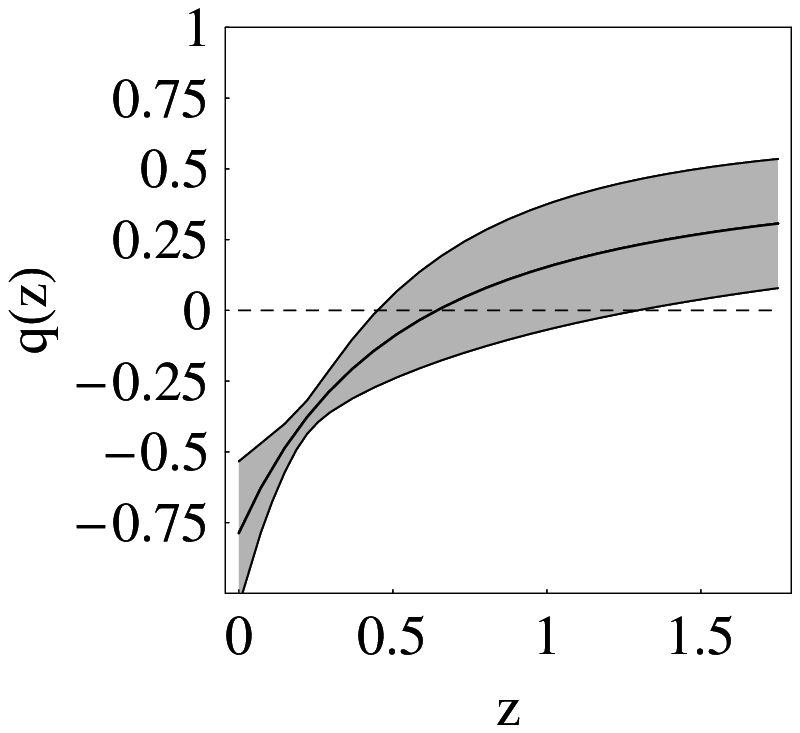}
  \includegraphics[width=6cm]{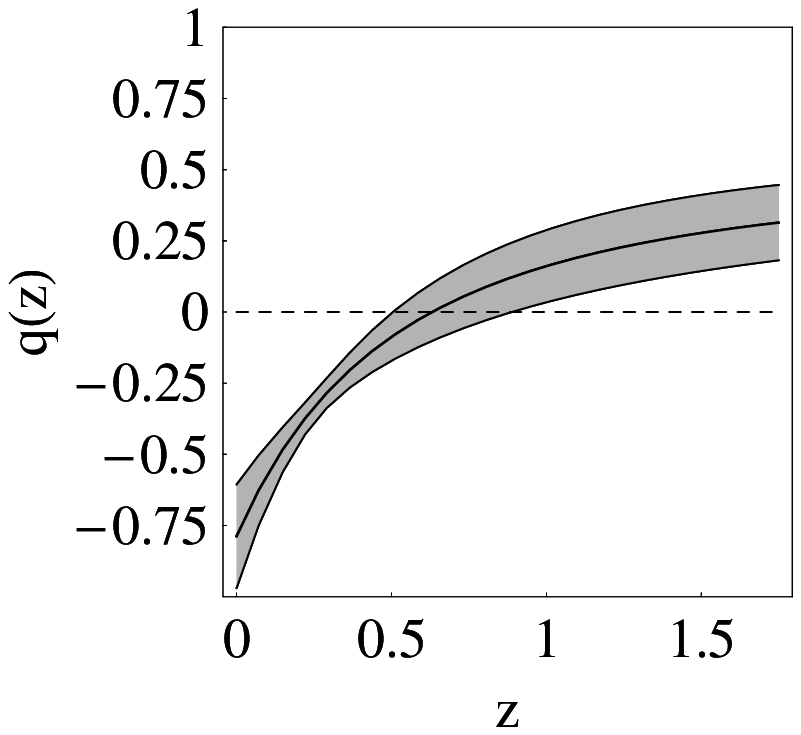}\\
~~~~~~~~~~~~~~~~~(a)~~~~~~~~~~~~~~~~~~~~~~~~~~~~~~~~~~~~~~~~~~~~~~~~~~~~~(b)\\
  \caption{The best fits of $q(z)$ with respect to redshift $z$ constrained from the 192 ESSENCE SNe Ia
  data
 (a) and its combination with
the 9 observational $H(z)$ data (b).}\label{fig:q1}
\end{figure}

 Thus on the basis of Eq. (\ref{5H2}), we can use the
maximum likelihood method for Eq. (\ref{1chi2-sne}) to constrain
parameters $(H_{0},a,b)$. It should be noticed that, since we are
interested in the model parameters $a, b$, the $H_{0}$ contained in
$\chi^{2}_{SNe}(H_{0},\theta)$ is a nuisance parameter and will be
marginalized by integrating
 the likelihood
$L(\theta) =\int d H_{0}P(H_{0})\exp$
 $(-\chi^{2}(H_{0},\theta)/2)$.
 $P(H_{0})$ is the prior distribution function of the current
Hubble constant, and a Gaussian prior $H_{0} = 72 \pm
8kmS^{-1}Mpc^{-1}$ \cite{24Freedman} is adopted in this paper. So,
by using the maximum likelihood method to minimize the quantity
$\chi^{2}_{SNe}$, we obtain the best fit model parameters
 $a=-1.287^{+0.381}_{-0.387}$,
$b=-0.099^{+1.727}_{-1.672}$ with $\chi_{min}^{2}=195.495$. It can
be seen that the number of degrees of freedom (dof) for this case is
190, here the value of dof of the model equals to the number of
observational data points
 minus the number of parameters. Then the
reduced $\chi^{2}$ value (i.e., the ratio of the $\chi^{2}_{min}$
value to the number of dof) is given as $\chi^{2}_{min}/$dof =
1.029. Furthermore, by fitting deceleration parameter
$q(z)=\frac{1}{2}+\frac{a+bz}{(1+z)^{2}}$ to the  192 ESSENCE SNe Ia
data,  the $1\sigma$ error of the best fit $q(z)$ calculated by
using the covariance matrix is plotted in FIG. \ref{fig:q1}(a). From
FIG. \ref{fig:q1}(a), it can be seen that the evolution of $q$ with
respect to redshift $z$ describes the current accelerating expansion
of universe and the decelerating expansion in the past, and the best
constraint on $q(z)$ from the 192 ESSENCE data lies in the redshift
range $z\sim 0.2-0.4$. Also, we can see that the best fit values of
transition redshift $z_{T}$ and current deceleration parameter
$q_{0}$ are $z_{T}=0.644^{+0.649}_{-0.194}$,
$q_{0}=-0.787^{+0.253}_{-0.253}$. We compare these results with the
ones obtained from Gold SNe Ia data. Considering Ref. \cite{35Gong},
where $z_{T}=0.36^{+0.24}_{-0.08}$ was obtained by using this
deceleration parameter to the 157 Gold data, it is shown that the
observations from 192 ESSENCE data tend to larger value of
transition redshift. According to Ref. \cite{0710.5690Ruth}, where
the model-independent method of using SNe Ia proposed and developed
by Daly and Djorgovski \cite{3Daly}\cite{5Daly} has been applied to
constrain  models of the accelerating universe from the 182 Gold SNe
Ia data. The results were given as $z_{T}=0.35^{+0.15}_{-0.07}$,
$q_{0}=-0.5^{+0.13}_{-0.13}$. We can see that this result of $z_{T}$
(or $q_{0}$) is smaller (or larger) than the 192 ESSENCE case.

In order to get the more stringent constraints on transition
redshift $z_{T}$ and current deceleration parameter $q_{0}$, we
combine the 192 ESSENCE data with the observational $H(z)$ data. The
Hubble parameter $H(z)$ depends on the differential age of the
universe as a function of redshift $z$ in the form
\begin{equation}
H(z)=-\frac{1}{1+z}\frac{dz}{dt}.
\end{equation}
Therefore, the value of $H(z)$ can be directly measured through a
determination of $dz/dt$. By using the differential ages of
passively evolving galaxies from the GDDS \cite{23Abraham} and
archival data \cite{24Treu}\cite{25Treu}, Simon $et$ $al$. obtained
nine values of $H(z)$ in the range of $0<z<1.8$
 \cite{22Simon}. Using these data we can constrain cosmological
models by minimizing
\begin{equation}
\chi^2_{Hub}(H_{0},
\theta)=\sum_{i=1}^{N}\frac{\left[H_{th}(z_i)-H_{obs}(H_{0},\theta,
z_i)\right]^2}{\sigma^2_{obs;i}},
\end{equation}
where $H_{th}$ is the predicted value for the Hubble parameter and
can be given by Eq. (\ref{5H2}), $H_{obs}$ is the observed value,
$\sigma^2_{obs;i}$ is the standard deviation measurement
uncertainty. Here the nuisance parameter $H_{0}$ is marginalized .
Then we combine  two datasets to minimize the total likelihood
$\chi^2_{total}$
\begin{equation}
\chi^2_{total}(a,b)=\chi^2_{SNe}(a,b)+\chi^2_{Hub}(a,b).\label{3chitotal}
\end{equation}

By using the maximum likelihood method for Eq. (\ref{3chitotal}), we
obtain the best fit model parameters  $a=-1.288^{+0.275}_{-0.276}$,
$b=-0.068^{+1.010}_{-0.998}$ with $\chi_{min}^{2}=205.254$. Here it
can be seen that dof = 199, and $\chi^{2}_{min}/$dof = 1.031. The
best fit evolution of $q(z)$   is plotted in FIG. \ref{fig:q1}(b)
for combined constraint from the 192 ESSENCE data and the 9
observational $H(z)$ data. From Fig. \ref{fig:q1}(b) we can see that
the best fit values of transition redshift
 $z_{T}$ and current deceleration parameter $q_{0}$  are
 $z_{T}=0.632^{+0.256}_{-0.127}$ and
$q_{0}=-0.788^{+0.182}_{-0.182}$. Replacing  the 192 ESSENCE data
with  the 182 Gold data in the combined constraint, we find that the
results for transition redshift and current deceleration parameter
are $z_{T}=0.502^{+0.180}_{-0.089}$,
$q_{0}=-0.692^{+0.202}_{-0.203}$. It is shown that the best fit
value of $z_{T}$ from the former combined constraint tends to larger
value than the latter one. The central value of $q_{0}$ from the
former combined constraint is smaller than the latter one. However,
at $1\sigma$ error range the value of $q_{0}$ is almost consistent
with being the same for the two  combined constraints.

\begin{normalsize}
\begin{center}
\bfseries 2.2 Constraining models of the accelerating universe using
model-independent EOS of dark energy
\end{center}
\end{normalsize}

Next we  use the model-independent EOS of dark energy to constrain
models of the accelerating universe and obtain the best fit values
of $z_{T}$ and $q_{0}$. To obtain significant constraints on
cosmological quantities, we combine other four cosmic observations:
the 3-year WMAP CMB shift parameter, the SDSS baryon acoustic peak,
the X-ray gas mass fraction in clusters, and the observational
$H(z)$ data from the GDDS and archival data with the two samples of
SNe Ia  to constrain DE models. And we compare the differences for
the constraints on cosmological quantities  $\Omega_{0m}$,
$w_{0de}$, $z_{T}$ and $q_{0}$ from the 192 ESSENCE data and the 182
Gold data with combining with other cosmic observations.

\begin{table}
\vspace*{-12pt}
\begin{center}
\begin{tabular}{|c | c |c| }
\hline Case model                  & $H(z)$ &$q(z)$
\\\hline
       $w_{de}(z)=w_{0}= $ const                  &$H_{0}(1+z)^{\frac{3}{2}}[\Omega_{0m}+(1-\Omega_{0m})(1+z)^{3w_{0}}]^{\frac{1}{2}}$ &
 $\frac{\Omega_{0m}+(1-\Omega_{0m})(1+3w_{0})(1+z)^{3w_{0}}}{2[\Omega_{0m}+(1-\Omega_{0m})(1+z)^{3w_{0}}]}$\\
       $w_{de}(z)=w_{0}+w_{1} \ln(1+z)$        & $H_{0}(1+z)^{\frac{3}{2}}[\Omega_{0m}+(1-\Omega_{0m})(1+z)^{3(w_{0}+w_{1})}]^{\frac{1}{2}}$ &
 $\frac{\Omega_{0m}+(1-\Omega_{0m})(1+z)^{3(w_{0}+w_{1}z)}[1+3w_{0}+3w_{1}+(1+z) \ln (1+z)^{3w_{1}}]}{2[\Omega_{0m}+(1-\Omega_{0m})(1+z)^{3(w_{0}+w_{1}z)}]}$ \\\hline
       \end{tabular}
       \end{center}
       \caption{The corresponding expressions of Hubble parameter $H(z)$ and
deceleration parameter $q(z)$ for two $w_{de}$ } \label{table1}
       \end{table}

 The structure of the anisotropies of
the cosmic microwave background radiation depends on two eras in
cosmology, i.e., last scattering and today. They can also be applied
to limit the model parameters of DE by using the shift parameter
\cite{25Bond}
\begin{equation}
R=\sqrt{\Omega_{0m}}\int_{0}^{z_{rec}}\frac{H_{0}dz^{'}}{H(z^{'};\theta)},\label{6R-CMB}
\end{equation}
where $z_{rec}=1089$ is the redshift of recombination. $R$ obtained
from the three-year WMAP data is \cite{26Spergel}
\begin{equation}
R=1.70\pm0.03.\label{7R-value}
\end{equation}
From the CMB constraint, the best fit values of parameters in the DE
models can be determined by minimizing
\begin{equation}
\chi^{2}_{CMB}(\theta)=\frac{(R(\theta)-1.70)^{2}}{0.03^{2}}.\label{8chi2-CMB}
\end{equation}

Because the universe has a fraction of baryons, the acoustic
oscillations in the relativistic plasma would be imprinted onto the
late-time power spectrum of the nonrelativistic matter
\cite{27Eisenstein}. Therefore, the acoustic signatures in the
large-scale clustering of galaxies can also serve as a test to
constrain models of DE with detection of a peak in the correlation
function of luminous red galaxies in the SDSS \cite{19Eisenstein}.
By using the equation
\begin{equation}
A=\sqrt{\Omega_{0m}}E(z_{BAO})^{-1/3}[\frac{1}{z_{BAO}}\int_{0}^{z}\frac{dz^{'}}{E(z^{'};\theta)}]^{2/3},\label{9A-BAO}
\end{equation}
and $A=0.469\pm0.017$ measured from the SDSS data, $z_{BAO}=0.35$,
 we can minimize the $\chi^{2}_{BAO}$defined as \cite{28Alam}
\begin{equation}
\chi^{2}_{BAO}(\theta)=\frac{(A(\theta)-0.469)^{2}}{0.017^{2}}.\label{10chi2-BAO}
\end{equation}
Where $E(z)$ is included in the Hubble parameter and can be given by
defining $H(z)=H_{0}E(z)$.

The observations of X-ray gas mass fraction in galaxy clusters
provide key information on the dark matter, on the formation of
structures in the universe, and can be used to constrain the
cosmological parameters \cite{29Arnaud}. It is assumed that the
baryon gas mass fraction in clusters \cite{30Nesseris}
\begin{equation}
f_{gas}=\frac{M_{b-gas}}{M_{tot}} \label{11f-gas}
\end{equation}
is constant, independent of redshift and is related to the global
fraction of the universe $\Omega_{b}/\Omega_{0m}$. In the standard
cold dark matter (SCDM) model, $f_{gas}^{SCDM}$ is \cite{30Nesseris}
\begin{equation}
f_{gas}^{SCDM}=\frac{b}{1+\alpha}\frac{\Omega_{b}}{\Omega_{0m}}(\frac{d_{A}^{SCDM}(z)}{d_{A}(z)})^{\frac{3}{2}},\label{12f-gasSCDM}
\end{equation}
where $d_{A}$ is diameter distance which relates with $d_{L}$ via
$d_{L}(z)=(1+z)^{2}d_{A}(z)$, the parameter $b$ is a bias factor
 suggesting that the baryon fraction in clusters is slightly lower than
for the universe as a whole, the parameter
$\alpha\simeq0.19\sqrt{h}$ is the ratio factor of optically luminous
baryonic mass with X-ray gas contained in clusters. From Cluster
Baryon Fraction (CBF), the best fit values of parameters in
cosmological model  can be determined by minimizing
\cite{30Nesseris}
\begin{equation}
\chi^{2}_{CBF}(\theta)=C-\frac{B^{2}}{A},\label{13chi2-CBF}
\end{equation}
where
\begin{eqnarray*}
A=\sum_{i=1}^{N}\frac{\widetilde{f}_{gas}^{SCDM}(z_{i})^{2}}{\sigma^{2}_{f_{gas,i}}},
\end{eqnarray*}

\begin{eqnarray*}
B=\sum_{i=1}^{N}\frac{\widetilde{f}_{gas}^{SCDM}(z_{i})\cdot
f_{gas,i}}{\sigma^{2}_{f_{gas,i}}},
\end{eqnarray*}

\begin{equation}
C=\sum_{i=1}^{N}\frac{f_{gas,i}^{2}}{\sigma^{2}_{f_{gas,i}}},\label{14ABC}
\end{equation}

and
\begin{equation}
\widetilde{f}_{gas}^{SCDM}(z_{i})=(\frac{d_{A}^{SCDM}(z)}{d_{A}(z)})^{\frac{3}{2}}.\label{15fgas}
\end{equation}
$N=26$ is the number of the observed $f_{gas,i}$ and
$\sigma^{2}_{gas,i}$ published in Ref. \cite{20Allen}.

\begin{table}
\vspace*{-12pt}
\begin{center}
\begin{tabular}{|c | c |c |c|c|}
\hline Data &Case model                                  &
$\chi^{2}_{min}$  &$\chi^{2}_{min}/$dof  & Best fit model parameters
\\\hline
           ESSENCE+$R$+$A$+$f_{gas}$+$H$ & $\Lambda$CDM               & 233.690  &1.034   &$\Omega_{0m}=0.264^{+0.017}_{-0.017}$ \\
           ~~~~~~~~ & $w_{de}(z)=w_{0}= $ const                 & 232.301        &1.032  & $\Omega_{0m}=0.263^{+0.027}_{-0.024}$, $w_{0}=-0.996^{+0.106}_{-0.116}$ \\
        ~~~~~~~~ & $w_{de}(z)=w_{0}+w_{1} \ln(1+z)$            & 231.106         &1.032  &$\Omega_{0m}=0.272^{+0.027}_{-0.024}$, $w_{0}=-1.041^{+0.127}_{-0.142}$,  $w_{1}=0.003^{+0.003}_{-0.084}$ \\
          Gold+$R$+$A$+$f_{gas}$+$H$ & $\Lambda$CDM         & 200.355       &0.928      & $\Omega_{0m}=0.280^{+0.019}_{-0.017}$  \\
       ~~~~~~~~ & $w_{de}(z)=w_{0}= $ const                        & 197.356 &0.918    &$\Omega_{0m}=0.280^{+0.028}_{-0.027}$, $w_{0}=-0.899^{+0.110}_{-0.122}$  \\
       ~~~~~~~~~ & $w_{de}(z)=w_{0}+w_{1} \ln(1+z)$             & 196.651  &0.918   &$\Omega_{0m}=0.287^{+0.028}_{-0.027}$, $w_{0}=-0.939^{+0.130}_{-0.149}$,  $w_{1}=0.002^{+0.003}_{-0.087}$ \\\hline
       \end{tabular}
       \end{center}
       \caption{The values of $\chi^{2}_{min}$, $\chi^{2}_{min}/$dof, and best fit model
parameters against the model} \label{table2}
       \end{table}

\begin{table}
\vspace*{-12pt}
\begin{center}
\begin{tabular}{|c | c| c| c|}
\hline Data &Case model                       & $z_{T}$ &$q_{0}$
\\\hline
          ESSENCE+$R$+$A$+$f_{gas}$+$H$  & $\Lambda$CDM              &$0.774^{+0.051}_{-0.050}$ &  $-0.605^{+0.025}_{-0.025} $\\
       ~~~~~~~~~~     & $w_{de}(z)=w_{0}= $ cosnt              &$0.776^{+0.055}_{-0.053}$& $-0.600^{+0.082}_{-0.083} $\\
       ~~~~~~~~~~    & $w_{de}(z)=w_{0}+w_{1} \ln(1+z)$    &$0.742^{+0.062}_{-0.056}$ &  $-0.637^{+0.091}_{-0.090} $\\
         Gold +$R$+$A$+$f_{gas}$+$H$   & $\Lambda$CDM              &$0.725^{+0.051}_{-0.051}$& $-0.579^{+0.026}_{-0.027}$ \\
       ~~~~~~~~~~    & $w_{de}(z)=w_{0}=$ cosnt              & $0.728^{+0.059}_{-0.061}$ &  $-0.471^{+0.088}_{-0.088}$ \\
      ~~~~~~~~~~`    &$w_{de}(z)=w_{0}+w_{1} \ln(1+z)$    &$0.706^{+0.063}_{-0.060}$ &$-0.504^{+0.097}_{-0.098} $ \\\hline
       \end{tabular}
       \end{center}
       \caption{The best fit values of transition redshift $z_{T}$
and current deceleration parameter $q_{0}$ against the model}
\label{table3}
       \end{table}

Next, using the datasets of above observational techniques, we
minimize the total likelihood $\chi^{2}_{total}$ \cite{FYWang}
\begin{equation}
\chi^{2}_{total}=\chi^{2}_{SNe}+\chi^{2}_{Hub}+\chi^{2}_{CMB}+\chi^{2}_{BAO}+\chi^{2}_{CBF}.\label{16chi2-total}
\end{equation}
In this paper we consider two combined constraints on DE models from
recently observed data, i.e., the 192 ESSENCE SNe Ia data and the
182 Gold SNe Ia data are combined with other four observational
datasets, respectively. $\chi^{2}_{total}$  for these two cases can
be written as $\chi^{2}_{total 1}=\chi^{2}_{192
SNe}+\chi^{2}_{Hub}+\chi^{2}_{CMB}+\chi^{2}_{BAO}+\chi^{2}_{CBF}$,
 and $\chi^{2}_{total 2}=\chi^{2}_{182
SNe}+\chi^{2}_{Hub}+\chi^{2}_{CMB}+\chi^{2}_{BAO}+\chi^{2}_{CBF}$.
For simplicity, we express the  cosmic observations as 192
ESSENCE+Hub+CMB+BAO+CBF and 182 Gold+Hub+CMB+BAO+CBF for the two
combined constraints in the following part.

Here we consider two model-independent EOS of DE, $w_{de}(z)=w_{0}=
$ const \cite{13Hannestad} (one-parameter model) and
$w_{de}(z)=w_{0}+w_{1}\ln(1+z)$ \cite{14Gerke} (two-parameter
model). For a flat universe, the corresponding Hubble parameter
$H(z)$ and deceleration parameter $q(z)$ are derived and listed in
TABLE \ref{table1} for these two $w_{de}(z)$. Besides, we also
consider the most popular  model $\Lambda$CDM. The Hubble parameter
$H(z)$ and deceleration parameter $q(z)$ for this model can be
obtained from TABLE \ref{table1} when $w_{0}=-1$ for the case of
$w_{de}(z)=w_{0}$.

Thus on the basis of the expressions of $H(z)$ and $q(z)$ in TABLE
\ref{table1}, we can obtain the best fit parameters against the
model  with its $\chi^{2}_{min}$ value  by using the maximum
likelihood method for Eq. (\ref{16chi2-total}). Furthermore,  the
reduced $\chi^{2}$  can also be calculated against the model.  TABLE
\ref{table2} lists the  results. From TABLE \ref{table2}  we can see
that the central value of current dimensionless matter density
$\Omega_{0m}$ is about $0.26 \sim0.27$ for the constraint from 192
ESSENCE+Hub+CMB+BAO+CBF. Comparing with the combined constraint from
182 Gold+Hub+CMB+BAO+CBF, it means the former combined constraint
tends to smaller current value of matter density $\Omega_{0m}$ than
the latter one, where the central value of $\Omega_{0m}$ is about
$0.28 \sim0.29$ for these three DE models. In addition, it is shown
that  for DE model
 $w_{de}(z)=w_{0}+w_{1}\ln(1+z)$, the best fit value of model
parameter $w_{1}$ has  small value for the both combined
constraints. It may be said that the model
$w_{de}(z)=w_{0}+w_{1}\ln(1+z)$ has a small correctional function
relative to the case of $w_{de}(z)=w_{0}=$ const. At last, it can be
seen that  for the case of $w_{de}(z)=w_{0}=$ const, the central
value of  current EOS of DE $w_{0de}$  is surprisingly close to
$\Lambda$CDM model ($w_{de}(z)=-1$) for the constraint from
 192 ESSENCE+Hub+CMB+BAO+CBF.

\begin{figure}[!htbp]
  \includegraphics[width=4cm]{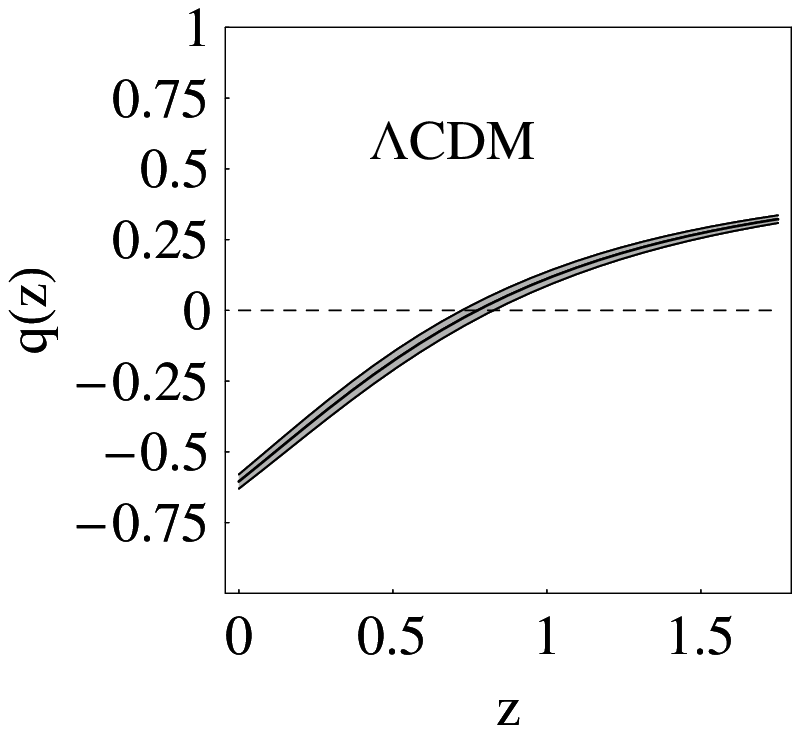}
 \includegraphics[width=4cm]{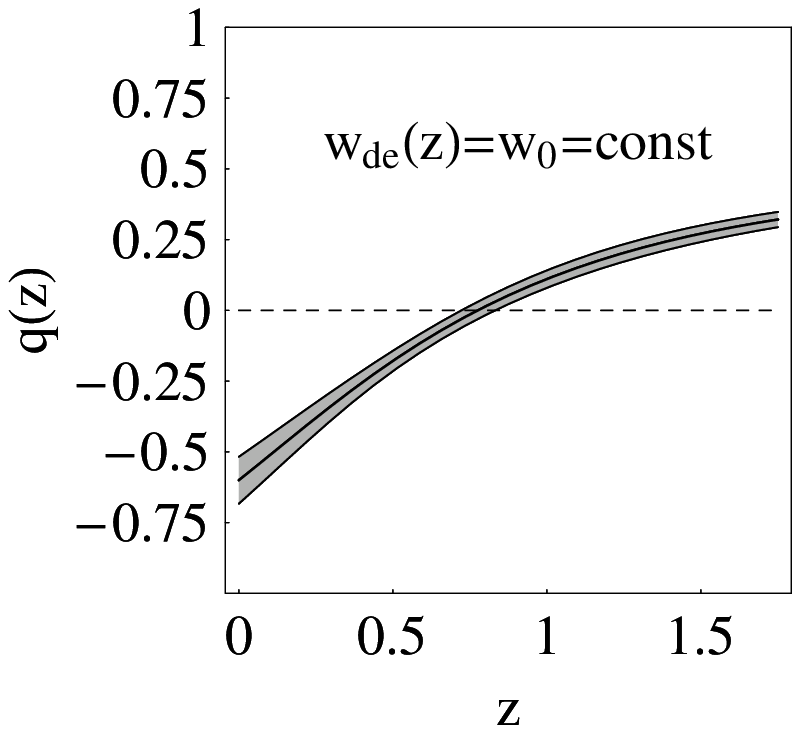}
  \includegraphics[width=4cm]{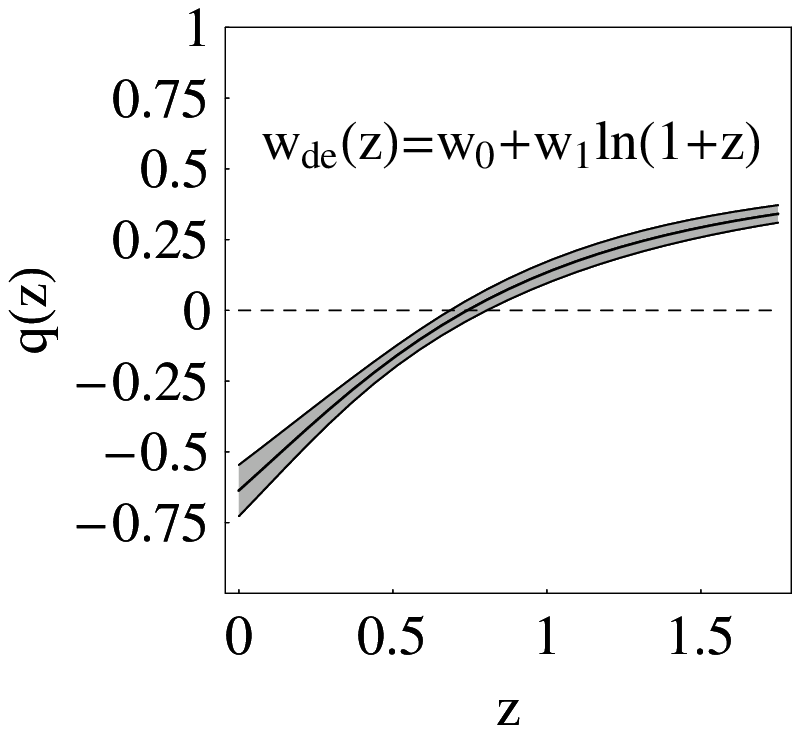}\\
  \includegraphics[width=4cm]{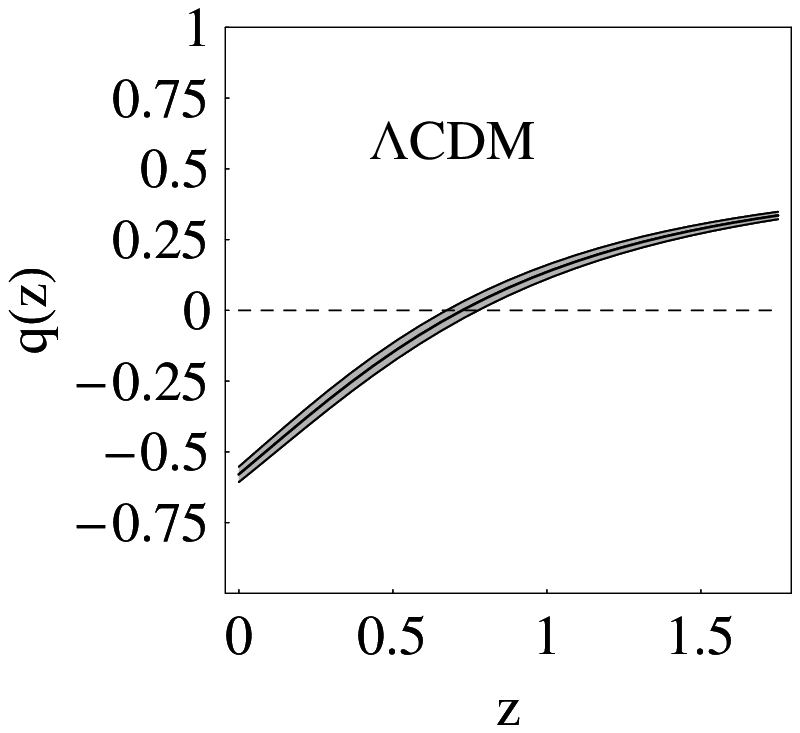}
 \includegraphics[width=4cm]{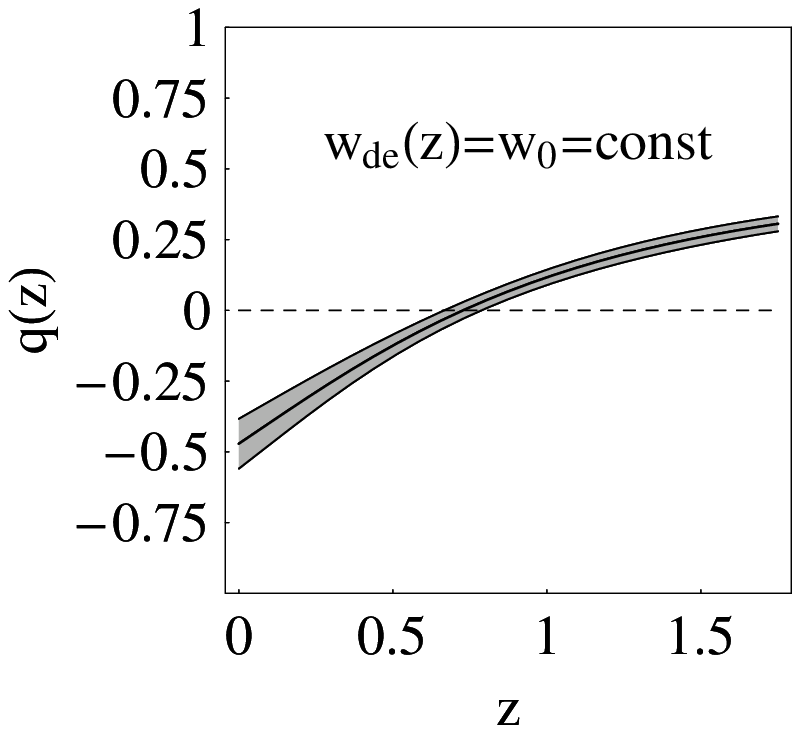}
 \includegraphics[width=4cm]{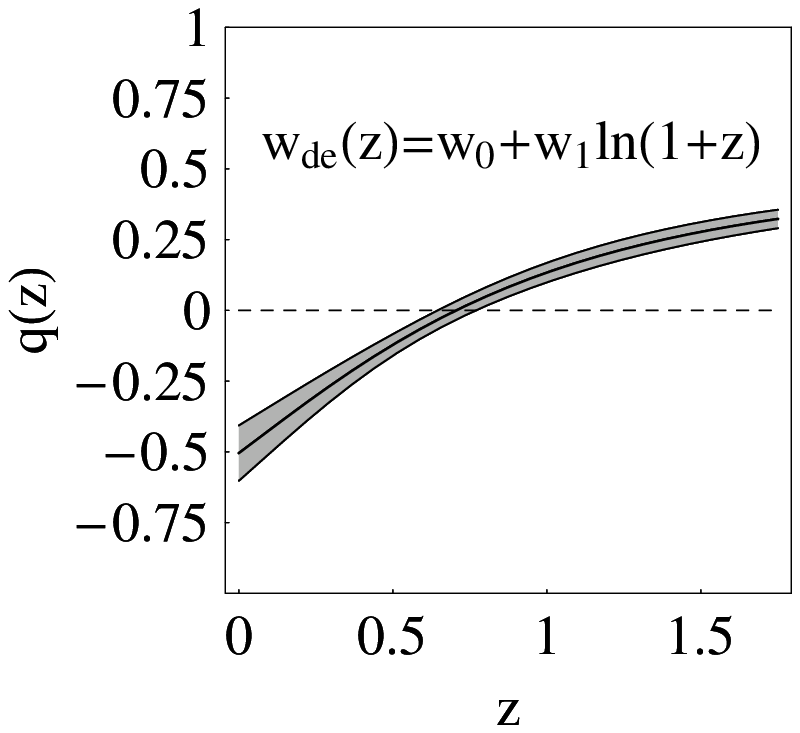}\\
  \caption{The best fits of $q(z)$
with $1\sigma$ error for three dark energy models constrained from
 192 ESSENCE+Hub+CMB+BAO+CBF (upper) and 182 Gold+Hub+CMB+BAO+CBF
(lower), respectively.}\label{fig:qw1}
\end{figure}

FIG.\ref{fig:qw1} shows the $1\sigma$ error of the best fit  $q(z)$
calculated by using the covariance matrix for three DE models. From
FIG.\ref{fig:qw1} we can get the best fit values of transition
redshift $z_{T}$ and current deceleration parameter $q_{0}$  against
the  model from two combined constraints. The results are listed in
TABLE \ref{table3}. We know that transition redshift $z_{T}$ denotes
the time when the evolution of universe changes from decelerated
expansion to accelerated expansion. The
 larger value of $z_{T}$, the earlier time of turning into an
accelerating universe. The value of $q_{0}$ indicates the expansion
rhythm of present universe. The  smaller value of $q_{0}$ , the more
violent of universe$^{,}$s acceleration. From TABLE \ref{table3}, it
can be found that, though at $1\sigma$ error range the differences
between the two combined constraints for the values of $z_{T}$ and
$q_{0}$ are not very obvious, the central values of $z_{T}$
constrained from 192 ESSENCE+Hub+CMB+BAO+CBF are bigger than the
cases  of 182 Gold+Hub+CMB+BAO+CBF, and the central values  of
$q_{0}$ are smaller than the cases  of 182 Gold SNe
Ia+Hub+CMB+BAO+CBF. Furthermore, we can see that the cosmic
acceleration could have started between the redshift
$z_{T}=0.706^{+0.063}_{-0.060}$ and $z_{T}=0.774^{+0.051}_{-0.050}$
for these two combined constraints.

\section{$\text{Model selection and Information criterion}$}
Since the emphasis of the ongoing and forthcoming research is
shifting from estimating specific parameters of the cosmological
model to model selection \cite{23Biesiada}, it is interesting to
estimate which model for an accelerating universe is distinguish by
statistical analysis of observational datasets out of a large number
of cosmological models. A popular but not too refined method to rate
goodness of models is to compare the quantity $\chi^{2}_{min}$/dof
\cite{07042606Lazkoz}. From TABLE \ref{table2} we can see that, both
the 192 ESSENCE and the 182 Gold data cases show a slightly higher
$\chi^{2}_{min}$/dof for the $\Lambda$CDM model, i.e.,  the
$\Lambda$CDM model has a less support from recent observations  when
compare it with other two DE models. In this paper we also use the
objective model selection criteria, including the Akaike information
criterion (AIC) and the Bayesian information criterion (BIC), to
estimate the strength of models.

In cosmology the information criterion (IC)  was first used by
Liddle \cite{37Liddle},  and then in subsequent papers
\cite{38Godlowski}\cite{39Szydlowski}.
 The AIC was derived by Akaike, and it takes the form
\begin{equation}
AIC=-2\ln {\cal L}(\theta \mid data)_{\max}+2K,\label{17}
\end{equation}
where ${\cal L}_{max}$ is the highest likelihood in the model with
the best fit parameters $\theta$, $K$ is the number of estimable
parameters ($\theta$) in the model. The term $-2\ln {\cal
L}(\theta\mid data)$ in Eq. (\ref{17}) is called $\chi^{2}$ and it
measures the quality of model fit, while the term $2K$ in Eq.
(\ref{17}) interprets model complexity. The BIC is similar to the
AIC, but the second term is different. It was derived by Schwarz and
is written as
\begin{equation}
BIC=-2\ln {\cal L}(\theta \mid data)_{\max}+K\ln n,
\end{equation}
where $n$ is the number of data points in the datasets.

\begin{table}
\vspace*{-12pt}
\begin{center}
\begin{tabular}{ | c| c| c|c|c|c|c|}
\hline Case model               &  AIC &$\Delta $
AIC$_{i}$ &$w_{i}$$_{(AIC)}$& BIC &$\Delta $ BIC$_{i}$ &$w_{i}$$_{(BIC)}$ \\
\hline
   $\Lambda$CDM                &  235.690  & 0       &  0.449     &239.115      & 0     & 0.712 \\
  $w_{de}(z)=w_{0}$=constant       & 236.301   &0.611   &  0.330          &243.151      & 4.036  &0.257 \\
 $w_{de}(z)=w_{0}+w_{1}\ln(1+z)$ & 237.106   & 1.416     &   0.221     &247.381      & 8.266  & 0.031     \\
\hline
\end{tabular}
       \end{center}
       \caption{The values of  AIC, AIC difference, AIC weight, BIC,  BIC difference and BIC weight  against the model for
        the constraint form 192 ESSENCE+Hub+CMB+BAO+CBF} \label{table4}
\end{table}

\begin{table}
\vspace*{-12pt}
\begin{center}
\begin{tabular}{ | c| c| c|c|c|c|c|}
\hline
 Case model                 &  AIC     &$\Delta $
AIC$_{i}$  &$w_{i}$$_{(AIC)}$   & BIC &$\Delta $ BIC$_{i}$
&$w_{i}$$_{(BIC)}$
\\ \hline
  $\Lambda$CDM              &  202.355  & 0.999        & 0.285     &205.735  & 0  &  0.525      \\
 $w_{de}(z)=w_{0}$=constant       & 201.356   &0       &   0.469      &208.116  & 2.381 & 0.434        \\
  $w_{de}(z)=w_{0}+w_{1}\ln(1+z)$ & 202.651   & 1.295    &    0.246     &212.791  & 7.056 &  0.041     \\
\hline
       \end{tabular}
       \end{center}
       \caption{The values of   AIC, AIC difference, AIC weight, BIC,  BIC difference and BIC weight against the model for
        the constraint form 182 Gold+Hub+CMB+BAO+CBF } \label{table5}
\end{table}

Now the question is how to assess the strength of models. We take
the AIC case as an example. The value of AIC  has no meaning by
itself for a single model and only the relative value between
different models are physically interesting. Therefore, by comparing
several models the one which minimizes the AIC is usually considered
the best, and denoted by AIC$_{min}$$=$min$\{$ AIC$_{i}$,
$i=1,...,N$$\}$, where $i=1,...,N$ is a set of alternative candidate
models. The relative strength of evidence for each model can be
obtained by calculating the relative likelihood of the model ${\cal
L}(M_{i}\mid data)\propto $ exp (-$\Delta$ AIC$_{i}$/2), where
$\Delta$ AIC$_{i}$= AIC$_{i}-$AIC$_{min}$  over the whole range of
alternative models. The Akaike weights $w_{i}$$_{(AIC)}$ are
calculated by normalizing the relative likelihoods of the models
${\cal L}(M_{i}\mid data)$ to unity. The rules for judging the AIC
model selection are as follows \cite{23Biesiada}: when $0\leq$
$\Delta$ AIC$_{i}$$\leq 2$
 model $i$ has almost the same support from the data
as the best model, for $2\leq $ $\Delta$ AIC$_{i}$$\leq 4$, model
$i$ is supported considerably less and with $\Delta$ AIC$_{i}$$ >10$
 model $i$ is practically irrelevant.  According to Eq. (\ref{17}) we
 can get the BIC values of several models. The model that have the minimum BIC value is
 considered the best. Then similar to the AIC case, taking it as a reference,  the BIC difference and BIC
 weight against the model
  can be  calculated. The rules for
judging the BIC model selection are described as
\cite{16Davis}\cite{37Liddle}: a $\Delta$BIC of more than 2 (or 6)
relative to the best one  is considered $^{"}$unsupported$^{"}$  (or
$^{"}$strongly unsupported$^{"}$) from observational data.
Furthermore, it should be noticed that according to Ref.
\cite{15Parkinson}, in the limit of large data points (large $n$)
AIC tends to favor models with more parameters while BIC tends to
penalize them. For more details about AIC and BIC, please see Refs.
\cite{23Biesiada}\cite{37Liddle}\cite{15Parkinson}\cite{38Godlowski}\cite{39Szydlowski}\cite{40Xu}\cite{41Szydlowski}.

In what follows, we will estimate which model is the best-fit one
according to the AIC and BIC for all the models in Table
\ref{table2}. Based on the values of $\chi^{2}_{min}$, it is shown
that the best model is the one following $\Lambda$CDM in terms of
its AIC value for the  constraint from 192 ESSENCE+Hub+CMB+BAO+CBF,
and the best one is $w_{de}(z)=w_{0}$ for the constraint from 182
Gold+Hub+CMB+BAO+CBF. For the case of each  combined constraint,
taking its minimum AIC value as a reference, we obtain the AIC
differences $\Delta$ AIC $_{i}$, Akaike weights $w_{i}$$_{(AIC)}$
against alternative models. TABLE \ref{table4} and \ref{table5} list
the calculating results for the  cases of two combined constraints.
Note that the model selection provides quantitative information to
judge the "strength of evidence", not just a way to select only one
model. From TABLE \ref{table4} and \ref{table5}, it can be seen that
three DE models have almost the same support from two datasets
because the values of $\Delta$ AIC$_{i}$ for other two models are in
the range 0-2 relative to the best one. The calculation for the BIC
is similar to the AIC case, and the results are listed in TABLE
\ref{table4} and \ref{table5}, too. In this analysis, we find the
best fit model is $\Lambda$CDM for the both combined constraints,
and the more free parameters in DE model, the weaker support from
observational data for these three DE models.

 According to the AIC  we can see that the
model degeneration is obvious  because  three DE models have almost
 the same support from observational data. Then we expect the new probers such as SNAP and Planck surveyor can
provide more accurate data and break up the model degeneration.
 For the BIC,  it is shown  that this model selection method can be a better one to avoid the model degeneration  than
the AIC.

\section{ $\text{Conclusion}$} \label{section3}

{~~~~In summary, we use the  192 ESSENCE SNe Ia data and the  182
Gold SNe Ia data combined with  other observed data such as the
3-year WMAP,
 the BAO peak from SDSS , the  X-ray gas
mass fraction in clusters  and the observational $H(z)$ data from
the GDDS and archival data,  to constrain models of the accelerating
universe.
 Concretely,  using the  192
  ESSENCE SNe Ia data and the 9 observational $H(z)$ data to
 constrain
a parameterized deceleration parameter
$q(z)=\frac{1}{2}+\frac{a+bz}{(1+z)^{2}}$
\cite{35Gong}\cite{15Gong},
 we obtain the best fit model parameters, $a=-1.288^{+0.275}_{-0.276}$,
$b=-0.068^{+1.010}_{-0.998}$. The best fit values of transition
redshift $z_{T}$ and current deceleration parameter $q_{0}$ are
given as $z_{T}=0.632^{+0.256}_{-0.127}$,
$q_{0}=-0.788^{+0.182}_{-0.182}$. Replacing the 192 ESSENCE data
with the 182 Gold data in combined constraint, it can be found that
at 1$\sigma$ error range, this result for $q_{0}$ is almost
consistent with being the same for the two combined constraints. But
 the result for $z_{T}$  at 1$\sigma$ error range tends to larger value than the case of the joint
analysis involving the 182 Gold data and the observational $H(z)$
data. Furthermore, it is shown that the central value of $z_{T}$ (or
$q_{0}$) for the former combined constraint is larger (or smaller)
than latter one. For producing the differences of these cosmological
quantities between these two combined constraints, the reason maybe
is caused by the different way that the SNe magnitudes are
calculated for the two samples of SNe Ia.  On the other hand, since
some data points in the two sets of SN data are from the different
subsets, some unknown system errors from the different instruments
for SNe surveys are also possible to contribute to these differences
for the cosmological quantities. We also expect the more advanced
probers to explore SNe in  future.

 Furthermore, for the cosmological quantities $\Omega_{0m}$, $w_{0de}$,
$z_{T}$ and $q_{0}$, we compare the differences for them between the
combined constraint  from the 192 ESSENCE data  with  other four
cosmic observations and the 182 Gold data with other four
observations. Considering DE model $\Lambda$CDM and
 two model-independent EOS of dark energy, $w_{de}(z)=w_{0}$,
$w_{de}(z)=w_{0}+w_{1}\ln(1+z)$,
 we plot the best fit forms of deceleration parameter $q(z)$
 with $1\sigma$ error by using
two sets of SNe data  with other cosmological observations. It can
be seen that the cosmic acceleration could have started between the
redshift $z_{T}=0.706^{+0.063}_{-0.060}$ and
$z_{T}=0.774^{+0.051}_{-0.050} $ for  two  combined constraints. By
comparing the two combined constraints on
 the DE models: $\Lambda$CDM, $w_{de}(z)=w_{0}$, and $w_{de}(z)=w_{0}+w_{1}\ln(1+z)$, we find that
the
 combined constraint from  192
ESSENCE+Hub+CMB+BAO+CBF tends to smaller current value of matter
density $\Omega_{0m}$ than the constraint from 182
Gold+Hub+CMB+BAO+CBF. And it can be seen that,
 though at $1\sigma$ error range the differences
between two combined constraints for the values of $z_{T}$ and
$q_{0}$ are not very obvious, the central values of $z_{T}$
constrained from 192 ESSENCE+Hub+CMB+BAO+CBF are bigger than the
cases  of 182 Gold+Hub+CMB+BAO+CBF, and the central values  of
$q_{0}$ are smaller than the cases  of 182 Gold SNe
Ia+Hub+CMB+BAO+CBF. At last, it is shown that for the case of
$w_{de}(z)=w_{0}=$ const, the central value of $w_{0de}$ constrained
from 192 ESSENCE+Hub+CMB+BAO+CBF is surprisingly close to
$\Lambda$CDM model ($w_{de}=-1$).

 Since it
is interesting to estimate which model for an accelerating universe
is distinguish by statistical analysis of observational datasets
over many models, by applying the recent observational data to the
objective information criterion  of model selection, we compare with
three DE models in TABLE \ref{table2} to assess the strength of
models. It is shown that, according to the AIC though the best model
is $\Lambda$CDM for using the combined datasets of 192
ESSENCE+Hub+CMB+BAO+CBF, other two models also have the same support
with the best one because the values of $\Delta$ AIC$_{i}$ for them
are in the range 0-2. For the case of the BIC we find the best fit
model is $\Lambda$CDM for both combined constraints, and the more
free parameters in DE model, the  weaker support from observational
data for these three DE models.

\textbf{\ Acknowledgments } The research work is supported by NSF
(10573003), NSF (10573004), NSF (10703001), and NSF (10647110), NBRP
(2003CB716300) of P.R. China.

\section{ $\text{Appendix}$} \label{section5}

\newcommand{\ZZ}[2]{\rule[#1]{0pt}{#2}}
\newcommand{\MC}[3]{\multicolumn{#1}{#2}{#3}}
\begin{center}\begin{tabular}{|c |c | c| c|c | c| c| c|c| c|c| }
\MC{11}{c}{\textbf {TABLE IV: ~The two sets of SNe Ia data with
their subsets.
}}\\
[5pt] \hline \MC{11}{|c|}{\ZZ{2pt}{2pt}\hfill\normalsize 192 ESSENCE
 SNe Ia data \hfill \vline\hfill\normalsize 182 Gold  SNe Ia data \hfill\mbox{}}\\
\hline \ZZ{2pt}{2pt}
 Number& SN &  $z$ &  $\mu$ &  $\sigma_{\mu}$ &Subsample &
SN & $z$ &  $\mu$ &  $\sigma_{\mu}$   &Subsample
\\\hline
1&b013   & 0.4260    &41.98      &0.23  & ESSENCE  & SN95K  & 0.478  & 42.48  & 0.23&HZSST\\
2 &d033   & 0.5310    & 42.96     &0.17 & ESSENCE  & SN96E   &0.425   &41.69   &0.40& HZSST \\
3 &d083   & 0.3330    &40.71      &0.14 &ESSENCE  & SN96H   &0.620   &43.11   &0.28& HZSST \\
4 &d084    &0.5190    &42.95      &0.29& ESSENCE  & SN96I  & 0.570  & 42.80   &0.25& HZSST \\
5 &d085    &0.4010    &41.96      &0.22& ESSENCE  & SN96J  & 0.300  & 41.01   &0.25& HZSST \\
6 &d086    &0.2050    &40.08      &0.30& ESSENCE  & SN96K   &0.380   &42.02  & 0.22& HZSST \\
7 &d089    &0.4360    &42.05      &0.20& ESSENCE  & SN96U  & 0.430  & 42.33   &0.34& HZSST \\
8 &d093    &0.3630    &41.73      &0.14& ESSENCE  & SN97as  & 0.508  & 42.19   &0.35& HZSST \\
9 &d097    &0.4360    &42.10      &0.17& ESSENCE  & SN97bb   &0.518   &42.83  & 0.31& HZSST \\
10 &d117    &0.3090    &41.42      &0.27& ESSENCE  & SN97bj  & 0.334   &40.92   &0.30& HZSST \\
11 &d149    &0.3420    &41.63      &0.21& ESSENCE  & SN97ce   &0.440   &42.07  & 0.19& HZSST \\
12 &e020    &0.1590    &39.79      &0.29& ESSENCE  & SN97cj   &0.500   &42.73  & 0.20& HZSST \\
13 &e029    &0.3320    &41.51      &0.28& ESSENCE  & SN98ac   &0.460   &41.81   &0.40& HZSST \\
14 &e108    &0.4690    &42.28      &0.16& ESSENCE  & SN98M   &0.630  & 43.26  & 0.37& HZSST \\
15 &e132    &0.2390    &40.42      &0.29& ESSENCE  &\ SN98J   &0.828   &43.59   &0.61& HZSST \\
16 &e136    &0.3520    &41.62      &0.27& ESSENCE  & SN99Q2$^{*}$
   &0.459  & 42.67   &0.22& HZSST \\
17 &e138    &0.6120    &42.99      &0.18& ESSENCE  & SN99U2   &0.511   &42.83  & 0.21& HZSST \\
18 &e140    &0.6310    &42.89      &0.18& ESSENCE  & SN99S$^{*}$  & 0.474   &42.81   &0.22& HZSST \\
19 &e147    &0.6450    &43.01      &0.18& ESSENCE  & SN99N   &0.537   &42.85  & 0.41& HZSST \\
20 &e148    &0.4290    &42.25      &0.20& ESSENCE  & SN99fn   &0.477  & 42.38  & 0.21& HZSST \\
21 &e149    &0.4970    &42.23      &0.26& ESSENCE  & SN99ff   &0.455   &42.29  & 0.28& HZSST \\
22 &f011    &0.5390    &42.66      &0.25& ESSENCE  & SN99fj  & 0.815   &43.75  & 0.33& HZSST \\
23 &f041    &0.5610    &42.72      &0.17& ESSENCE  & SN99fm  & 0.949  & 44.00   &0.24& HZSST \\
24 &f231    &0.6190    &43.05      &0.17& ESSENCE  & SN99fk   &1.056  & 44.35  & 0.23& HZSST \\
25 &f235    &0.4220    &41.78      &0.24& ESSENCE  & SN99fw  & 0.278   &41.01   &0.41 & HZSST \\
26 &f244    &0.5400    &42.72      &0.26& ESSENCE  & SN99fv$^{*}$   &1.199  & 44.19  & 0.34& HZSST \\
27 &g005    &0.2180    &40.37      &0.26& ESSENCE  & SN00ec$^{*}$  & 0.470   &42.76  & 0.21& HZSST \\
28 &g050    &0.6330    &42.77      &0.18& ESSENCE  & SN00dz   &0.500   &42.74  & 0.24& HZSST \\
29 &g052    &0.3830    &41.56      &0.22& ESSENCE  & SN00eg  & 0.540  & 41.96   &0.41& HZSST \\
30 &g055    &0.3020    &41.39      &0.37& ESSENCE  & SN00ee$^{*}$   &0.470  & 42.73  & 0.23& HZSST \\
31 &g097    &0.3400    &41.56      &0.31& ESSENCE  & SN00eh  & 0.490  & 42.40  & 0.25& HZSST \\
32 &g120    &0.5100    &42.30      &0.21& ESSENCE  & SN01jh  & 0.884  & 44.22  & 0.19& HZSST \\
33 &g133    &0.4210    &42.22      &0.33& ESSENCE  & SN01hu  & 0.882   &43.89   &0.30& HZSST \\
34 &g142    &0.3990    &41.96      &0.43& ESSENCE  & SN01iy   &0.570   &42.87  & 0.31& HZSST \\
35 &g160    &0.4930    &42.38      &0.26& ESSENCE  & SN01jp   &0.528   &42.76  & 0.25& HZSST \\
36 &g240    &0.6870    &43.04      &0.20& ESSENCE  & SN01fo$^{*}$  & 0.771  & 43.12   &0.17& HZSST \\
37 &h283    &0.5020    &42.49      &0.37& ESSENCE  & SN01hs  & 0.832   &43.55   &0.29& HZSST \\
38 &h300    &0.6870    &43.09      &0.17& ESSENCE  & SN01hx  & 0.798  & 43.88  & 0.31& HZSST \\
39 &h319    &0.4950    &42.40      &0.21& ESSENCE  & SN01hy  & 0.811  & 43.97  & 0.35 & HZSST \\
40 &h323    &0.6030    &43.01      &0.22& ESSENCE  & SN01jf  & 0.815   &44.09   &0.28 & HZSST \\
\hline
\end{tabular}\end{center}

\newcommand{\YY}[2]{\rule[#1]{0pt}{#2}}
\newcommand{\MD}[3]{\multicolumn{#1}{#2}{#3}}
\begin{center}\begin{tabular}{|c |c | c| c|c | c| c| c|c| c|c| }
\MD{11}{c}{\textbf {~~~~~~TABLE IV continued~~ }}\\
[5pt] \hline \MD{11}{|c|}{\YY{2pt}{2pt}\hfill\normalsize ~~~~~192
ESSENCE  SNe Ia data \hfill \vline\hfill\normalsize 182 Gold SNe Ia  data \hfill\mbox{}}\\
\hline \YY{2pt}{2pt}
 Number& SN &  $z$ &  $\mu$ &  $\sigma_{\mu}$ &Subsample &
SN & $z$ &  $\mu$ &  $\sigma_{\mu}$   &Subsample
\\\hline
41 &h342    &0.4210    &42.18      &0.16& ESSENCE  & SN01jm   &0.977   &43.91  & 0.26& HZSST \\
42 &h359    &0.3480    &41.89      &0.27& ESSENCE  & SN95aw  & 0.400   &42.04   &0.19& SCP \\
43 &h363    &0.2130    &40.33      &0.33& ESSENCE  & SN95ax   &0.615   &42.85  & 0.23& SCP \\
44 &h364    &0.3440    &41.32      &0.17& ESSENCE  & SN95ay &0.480
&42.37  & 0.20& SCP \\
45 &k425    &0.2740    &41.12      &0.28& ESSENCE  & SN95az  & 0.450  & 42.13   &0.21& SCP \\
46 &k429    &0.1810    &39.89      &0.17& ESSENCE  & SN95ba   &0.388   &42.07  & 0.19& SCP \\
47 &k448    &0.4010    &42.34      &0.40& ESSENCE  & SN96ci  & 0.495   &42.25   &0.19& SCP \\
48 &k485    &0.4160    &42.16      &0.39& ESSENCE  & SN96cl   &0.828   &43.96  & 0.46& SCP \\
49 &m027    &0.2860    &41.53      &0.32& ESSENCE  & SN97eq   &0.538   &42.66   &0.18& SCP \\
50 &m158    &0.4630    &42.58      &0.28& ESSENCE  & SN97ek   &0.860   &44.03  & 0.30& SCP \\
51 &m193    &0.3410    &41.29      &0.23& ESSENCE  & SN97ez   &0.778  & 43.81  & 0.35& SCP \\
52 &n256    &0.6310    &43.09      &0.15& ESSENCE  & SN97F   &0.580  & 43.04   &0.21& SCP \\
53 &n263    &0.3680    &41.56      &0.17& ESSENCE  & SN97H  & 0.526  & 42.56  & 0.18& SCP \\
54 &n278    &0.3090    &41.16      &0.21& ESSENCE  & SN97I  & 0.172  & 39.79  & 0.18& SCP \\
55 &n285    &0.5280    &42.63      &0.26& ESSENCE  & SN97N   &0.180   &39.98   &0.18& SCP \\
56 &n326    &0.2680    &40.81      &0.26& ESSENCE  & SN97P   &0.472   &42.46  & 0.19& SCP \\
57 &n404    &0.2160    &40.59      &0.31& ESSENCE  & SN97Q  & 0.430   &41.99   &0.18& SCP \\
58 &p454    &0.6950    &43.53      &0.17& ESSENCE  & SN97R   &0.657   &43.27  & 0.20& SCP \\
59 &p455    &0.2840    &41.10      &0.29& ESSENCE  & SN97ac   &0.320   &41.45  & 0.18& SCP \\
60 &p524    &0.5080    &42.43      &0.22& ESSENCE  & SN97af   &0.579  & 42.86   &0.19& SCP \\
61 &SN92ag    &0.0259    &35.14      &0.22 & nearby  & SN97ai  & 0.450   &42.10   &0.23& SCP \\
62 &SN92bc    &0.0198    &34.84      &0.18 & nearby  & SN97aj   &0.581   &42.63   &0.19& SCP \\
63 &SN92bo    &0.0181    &34.73      &0.21 & nearby  & SN97am   &0.416   &42.10  & 0.19& SCP \\
64 &SN94S    &0.0160    &34.35      &0.22 & nearby  & SN97ap   &0.830  & 43.85   &0.19& SCP \\
65 &SN95ak    &0.0220    &34.70      &0.21 & nearby  & SN98ba  & 0.430   &42.36  & 0.25& SCP \\
66 &SN96bo    &0.0163    &33.98      &0.24 &nearby  & SN98bi  & 0.740   &43.35   &0.30& SCP \\
67 &SN97Y    &0.0166    &34.53      &0.23 & nearby  & SN00fr   &0.543   &42.67  & 0.19&SCP\\
68 &SN98V    &0.0172    &34.36      &0.22 & nearby  & SN90T  & 0.040   &  36.38 &  0.20 & LR\\
69 &SN98ab    &0.0279    &35.17      &0.18 & nearby  & SN91U  & 0.033   &35.53  & 0.21 & LR\\
70 &SN98ef    &0.0167    &34.16      &0.23 & nearby  & SN91S   &0.056   &37.31   &0.19 & LR\\
71 &SN98eg    &0.0235    &35.32      &0.20 & nearby  & SN92bg   &0.036   &36.17  & 0.20& LR\\
72 &SN99aw    &0.0392    &36.54      &0.13 & nearby  & SN92bk   &0.058  & 37.13   &0.19& LR\\
73 &SN99ek    &0.0176    &34.28      &0.22 & nearby  & SN92J   &0.046   &36.35   &0.21 & LR\\
74 &SN00ca    &0.0245    &35.24      &0.17 & nearby  & SN92au   &0.061   &37.31  & 0.22& LR\\
75 &SN00cn    &0.0232    &35.12      &0.18 & nearby  & SN93ah   &0.028   &35.53  & 0.22& LR\\
76 &SN00dk    &0.0164    &34.37      &0.22 & nearby  & SN94Q   &0.029   &35.70  & 0.21& LR\\
77 &SN00fa    &0.0218    &34.90      &0.21 & nearby  & SN98cs  & 0.032   &36.08  & 0.20& LR\\
78 &SN01V    &0.0162    &34.14      &0.22 & nearby  & SN99ef  & 0.038   &36.67   &0.19& LR\\
79 &SN01ba    &0.0305    &35.88      &0.16 & nearby  & SN99X  & 0.025  & 35.40  & 0.22 & LR\\
80 &SN01cz    &0.0163    &34.28      &0.24 & nearby  & SN00bk  & 0.026   &35.35  & 0.23& LR\\
\hline
\end{tabular}\end{center}

\newcommand{\XX}[2]{\rule[#1]{0pt}{#2}}
\newcommand{\ME}[3]{\multicolumn{#1}{#2}{#3}}
\begin{center}\begin{tabular}{|c |c | c| c|c | c| c| c|c| c|c| }
\ME{11}{c}{\textbf {~TABLE IV continued }}\\
[5pt] \hline \ME{11}{|c|}{\XX{2pt}{2pt}\hfill\normalsize ~~~~~~192
ESSENCE  SNe Ia data \hfill \vline\hfill\normalsize 182 Gold  SNe Ia data \hfill\mbox{}}\\
\hline \XX{2pt}{2pt}
 Number& SN &  $z$ &  $\mu$ &  $\sigma_{\mu}$ &Subsample &
SN & $z$ &  $\mu$ &  $\sigma_{\mu}$   &Subsample
\\\hline
81 &SN90O    &0.0306    &35.81      &0.17  & nearby  & SN90O  & 0.030  & 35.90  & 0.21& LR\\
82 &SN90af    &0.0502    &36.69      &0.20 & nearby  & SN90af   &0.050  & 36.84  & 0.22& LR\\
83 &SN92P    &0.0263    &35.60      &0.19 & nearby  & SN92P  & 0.026  & 35.63  & 0.22& LR\\
84 &SN92ae    &0.0748    &37.72      &0.21 & nearby  & SN92ae  & 0.075   &37.77   &0.19& LR\\
85 &SN92aq    &0.1009    &38.80      &0.15 & nearby  & SN92aq  & 0.101  & 38.70  &0.20& LR\\
86 &SN92bh    &0.0451    &36.91      &0.19 & nearby  & SN92bh  & 0.045  & 36.99  & 0.18& LR\\
87 &SN92bl    &0.0429    &36.49      &0.18 & nearby  & SN92bl  & 0.043   &36.52  & 0.19& LR\\
88 &SN92bp    &0.0789    &37.78      &0.15 & nearby  & SN92bp  &
0.079   &37.94   &0.18& LR\\
89 &SN92br    &0.0878    &37.76      &0.23 & nearby  & SN92br  & 0.088  & 38.07  & 0.28& LR\\
90 &SN92bs    &0.0634    &37.64      &0.20 &nearby  & SN92bs  & 0.063  & 37.67  & 0.19& LR\\
91 &SN93B    &0.0707    &37.78      &0.19 & nearby  & SN93B  & 0.071  & 37.78  & 0.19& LR\\
92 &SN93H    &0.0248    &35.10      &0.18 & nearby  & SN93H & 0.025  & 35.09  & 0.22& LR\\
93 &SN93O    &0.0519    &37.12      &0.15 & nearby  & SN93O  & 0.052  & 37.16   &0.18& LR\\
94&SN93ag    &0.0500    &37.07      &0.18 & nearby  & SN93ag  & 0.050  & 37.07  & 0.19& LR\\
95& SN94M    &0.0243    &35.24      &0.20 & nearby  & SN94M  & 0.024  & 35.09  & 0.22& LR\\
96& SN94T    &0.0357    &36.02      &0.17 & nearby  & SN94T  & 0.036   &36.01   &0.21& LR\\
97& SN95ac    &0.0488    &36.57     &0.16 & nearby  & SN95ac  & 0.049  & 36.55  & 0.20& LR\\
98& SN96C    &0.0275    &35.94      &0.18 & nearby  & SN96C  & 0.027  & 35.90  & 0.21& LR\\
99& SN96ab    &0.1242    &38.90      &0.20 & nearby  & SN96ab  & 0.124   &39.19  & 0.22& LR\\
100& SN96bl    &0.0348    &36.09      &0.18 & nearby  & SN96bl  & 0.034   &36.19  & 0.20& LR\\
101& SN97dg    &0.0297    &36.15      &0.19 & nearby  & SN97dg  & 0.029   &36.13  & 0.21& LR\\
102& SN98dx    &0.0537    &36.92      &0.15 & nearby  & SN98dx  & 0.053  & 36.95  & 0.19& LR\\
103& SN99cc    &0.0315    &35.82      &0.17 & nearby  & SN99cc &  0.031  & 35.84  & 0.21& LR\\
104& SN99gp    &0.0260    &35.62      &0.16 & nearby  & SN99gp  & 0.026 &  35.57  & 0.21& LR\\
105& SN00cf    &0.0365    &36.36      &0.17 & nearby  & SN00cf &  0.036  & 36.39  & 0.19& LR\\
106& 1977ff    &1.7550    &45.31      &0.36 & HST  & 1997ff  & 1.755   &45.35  & 0.35& HST\\
107& 2002dc    &0.4750    &42.20      &0.21 & HST  & 2002dc  & 0.475   &42.24  & 0.20& HST\\
108& 2002dd    &0.9500    &43.94      &0.35 & HST  & 2002dd  & 0.950  & 43.98   &0.34& HST\\
109& 2003eq    &0.8400    &43.63      &0.22 & HST  & 2003eq  & 0.840  & 43.67   &0.21& HST\\
110& 2003es    &0.9540    &44.26      &0.28 & HST  & 2003es   &0.954   &44.30   &0.27& HST\\
111& 2003eb    &0.9000    &43.60      &0.26 & HST  &  2003eb   &0.900  & 43.64  & 0.25& HST\\
112& 2003XX    &0.9350    &43.93      &0.30 & HST  &  2003XX  & 0.935   &43.97  & 0.29& HST\\
113& 2003bd    &0.6700    &43.15      &0.25 & HST  & 2003bd   &0.670  & 43.19  & 0.24& HST\\
114& 2002kd    &0.7350    &43.10      &0.20 & HST  & 2002kd   &0.735   &43.14  & 0.19& HST\\
115& 2003be    &0.6400    &42.97      &0.26 & HST  & 2003be   &0.640   &43.01   &0.25& HST\\
116& 2003dy    &1.3400    &44.88      &0.32 & HST  &  2003dy  & 1.340  & 44.92   &0.31& HST\\
117& 2002ki    &1.1400    &44.67      &0.30 & HST  & 2002ki  & 1.140  & 44.71   &0.29& HST\\
118& 2002hp    &1.3050    &44.47      &0.31 & HST  & 2002hp   &1.305  & 44.51  & 0.30& HST\\
119& 2002fw    &1.3000    &45.02      &0.21 & HST  &  2002fw   &1.300   &45.06  & 0.20& HST\\
120& HST04Pat    &0.9700    &44.63      &0.37 & HST  & HST04Pat   &0.970   &44.67  & 0.36& HST\\
\hline
\end{tabular}\end{center}

\newcommand{\WW}[2]{\rule[#1]{0pt}{#2}}
\newcommand{\MF}[3]{\multicolumn{#1}{#2}{#3}}
\begin{center}\begin{tabular}{|c |c | c| c|c | c| c| c|c| c|c| }
\MF{11}{c}{\textbf {TABLE IV continued }}\\
[5pt] \hline \MF{11}{|c|}{\WW{2pt}{2pt}\hfill\normalsize ~~~~~~192
ESSENCE  SNe Ia data \hfill \vline\hfill\normalsize 182 Gold  SNe Ia data \hfill\mbox{}}\\
\hline \WW{2pt}{2pt}
 Number& SN &  $z$ &  $\mu$ &  $\sigma_{\mu}$ &Subsample &
SN & $z$ &  $\mu$ &  $\sigma_{\mu}$   &Subsample
\\\hline
121& HST04Mcg    &1.3700    &45.19      &0.26 & HST  & HST04Mcg  & 1.370  & 45.23  & 0.25& HST\\
122& HST05Fer    &1.0200    &43.95      &0.28 & HST  & HST05Fer  & 1.020  & 43.99  & 0.27& HST\\
123& HST05Koe    &1.2300    &45.13      &0.24 & HST  & HST05Koe   &1.230  & 45.17   &0.23& HST\\
124& HST04Gre    &1.1400    &44.40      &0.32 & HST  & HST04Gre  & 1.140  & 44.44   &0.31& HST\\
125& HST04Omb    &0.9750    &44.17      &0.27 & HST  & HST04Omb   &0.975   &44.21  & 0.26& HST\\
126& HST05Lan    &1.2300    &44.93      &0.21 & HST  & HST05Lan  & 1.230  & 44.97   &0.20& HST\\
127& HST04Tha    &0.9540    &43.81      &0.28 & HST  & HST04Tha   &0.954   &43.85  & 0.27& HST\\
128& HST04Rak    &0.7400    &43.34      &0.23 & HST  & HST04Rak  & 0.740   &43.38  & 0.22& HST\\
129& HST04Yow    &0.4600    &42.19      &0.33 & HST  & HST04Yow  & 0.460  & 42.23 &  0.32& HST\\
130& HST04Man    &0.8540    &43.92      &0.30 & HST  & HST04Man  & 0.854  & 43.96  & 0.29& HST\\
131& HST05Spo    &0.8390    &43.41      &0.21 & HST  & HST05Spo   &0.839   &43.45  & 0.20& HST\\
132& HST04Eag    &1.0200    &44.48      &0.20 & HST  & HST04Eag   &1.020  & 44.52  & 0.19& HST\\
133& HST05Gab    &1.1200    &44.63      &0.19 & HST  & HST05Gab  & 1.120  & 44.67  & 0.18& HST\\
134& HST05Str    &1.0100    &44.73      &0.20 & HST  & HST05Str   &1.010   &44.77   &0.19& HST\\
135& HST04Sas    &1.3900    &44.86      &0.20 & HST  & HST04Sas   &1.390  & 44.90  & 0.19 & HST\\
136& SN03D1au    &0.5043    &42.55      &0.18  &SNLS & SN03D1au  & 0.504   & 42.61   &0.17&SNLS\\
137& SN03D1aw    &0.5817    &43.12      &0.21 &SNLS & SN03D1aw   &0.582   & 43.07   & 0.17&SNLS\\
138& SN03D1ax    &0.4960    &42.33      &0.20 &SNLS & SN03D1ax   &0.496   & 42.36    &0.17&SNLS\\
139& SN03D1co    &0.6790    &43.59      &0.27 &SNLS & SN03D1co   &0.679   & 43.58   & 0.19&SNLS\\
140& SN03D1fc    &0.3310    &41.30      &0.21 &SNLS & SN03D1fc   &0.331    &41.13    &0.17&SNLS\\
 141& SN03D1fl    &0.6880    &43.13      &0.23 &SNLS & SN03D1fl   &0.688    &43.23   & 0.17&SNLS\\
142& SN03D1fq    &0.8000    &43.92      &0.27 &SNLS & SN03D1fq   &0.800    &43.67   & 0.19&SNLS\\
143& SN03D3af    &0.5320    &42.84      &0.29 &SNLS & SN03D3af   &0.532   & 42.78   & 0.18&SNLS\\
144& SN03D3aw    &0.4490    &42.07      &0.24 &SNLS & SN03D3aw   &0.449  &42.05    &0.17&SNLS\\
145& SN03D3ay    &0.3709    &41.80      &0.23 &SNLS & SN03D3ay   &0.371   & 41.67   & 0.17&SNLS\\
146& SN03D3cc    &0.4627    &42.25      &0.17 &SNLS & SN03D3cc   &0.463   & 42.27   & 0.17&SNLS\\
147& SN03D3cd    &0.4607    &42.12      &0.18 &SNLS & SN03D3cd   &0.461   & 42.22    &0.17&SNLS\\
148& SN03D4ag    &0.2850    &40.98      &0.13 &SNLS & SN03D4ag   &0.285    &40.92  &0.17&SNLS\\
149& SN03D4at    &0.6330    &43.26      &0.25 &SNLS & SN03D4at   &0.633   & 43.32    &0.18&SNLS\\
150& SN03D4cx    &0.9490    &44.26      &0.17 &SNLS & SN03D4cx   &0.949   & 43.69   & 0.32&SNLS\\
151& SN03D4cz    &0.6950    &43.11      &0.34 &SNLS & SN03D4cz   &0.695   & 43.21    &0.19&SNLS\\
152& SN03D4dh    &0.6268    &42.94      &0.23 &SNLS & SN03D4dh   &0.627    &42.93    &0.17&SNLS\\
153& SN03D4di    &0.9050    &43.84      &0.16 &SNLS & SN03D4di   &0.905   &43.89  &0.30&SNLS\\
154& SN03D4dy    &0.6040    &42.79      &0.32 &SNLS & SN03D4dy   & 0.604  &42.70    &0.17&SNLS\\
155& SN03D4fd    &0.7910    &43.75      &0.21 &SNLS & SN03D4fd   &0.791  &43.54   & 0.18&SNLS\\
156& SN03D4gg    &0.5920    &42.87      &0.25 &SNLS & SN03D4gg   &0.592   & 42.75    &0.19&SNLS\\
157& SN04D2fp    &0.4150    &42.06      &0.17 &SNLS & SN04D2fp   &0.415   & 41.96    &0.17&SNLS\\
158& SN04D2fs    &0.3570    &41.65      &0.22 &SNLS & SN04D2fs   &0.357   & 41.63   & 0.17&SNLS\\
159& SN04D2gb    &0.4300    &41.81      &0.18 &SNLS & SN04D2gb   &0.430   & 41.96   & 0.17&SNLS\\
160& SN04D3co    &0.6200    &43.20      &0.26 &SNLS & SN04D3co   &0.620   & 43.21   & 0.18&SNLS\\
\hline
\end{tabular}\end{center}

\newcommand{\VV}[2]{\rule[#1]{0pt}{#2}}
\newcommand{\MG}[3]{\multicolumn{#1}{#2}{#3}}
\begin{center}\begin{tabular}{|c |c | c| c|c | c| c| c|c| c|c| }
\MG{11}{c}{\textbf {TABLE IV continued }}\\
[5pt] \hline \MG{11}{|c|}{\VV{2pt}{2pt}\hfill\normalsize ~~~~~192
ESSENCE  SNe Ia data \hfill \vline\hfill\normalsize 182 Gold SNe Ia  data \hfill\mbox{}}\\
\hline \VV{2pt}{2pt}
 Number& SN &  $z$ &  $\mu$ &  $\sigma_{\mu}$ &Subsample &
SN & $z$ &  $\mu$ &  $\sigma_{\mu}$   &Subsample
\\\hline
161& SN04D3cy    &0.6430    &43.34      &0.22 &SNLS & SN04D3cy   &0.643   &43.21    &0.18 &SNLS\\
162& SN04D3df    &0.4700    &42.03      &0.20 &SNLS & SN04D3df   &0.470   &42.45   & 0.17 &SNLS\\
163& SN04D3do    &0.6100    &42.82      &0.29 &SNLS & SN04D3do   &0.610    &42.98   & 0.17 &SNLS\\
164& SN04D3ez    &0.2630    &40.76      &0.21 &SNLS & SN04D3ez   &0.263   &40.87   & 0.17 &SNLS\\
165& SN04D3fk    &0.3578    &41.41      &0.21 &SNLS & SN04D3fk   &0.358   & 41.66   & 0.17 &SNLS\\
166& SN04D3fq    &0.7300    &43.57      &0.26 &SNLS & SN04D3fq   &0.730   & 43.47   & 0.18 &SNLS\\
167& SN04D3hn    &0.5516    &42.28      &0.41 &SNLS & SN04D3hn   &0.552   &42.65   &0.17 &SNLS\\
168& SN04D3kr    &0.3373    &41.46      &0.17 &SNLS & SN04D3kr   &0.337   &41.44   &0.17 &SNLS\\
169& SN04D3lu    &0.8218    &43.76      &0.22 &SNLS & SN04D3lu   &0.822   &43.73   & 0.27 &SNLS\\
170& SN04D3ml    &0.9500    &44.14      &0.15 &SNLS & SN04D3ml   &0.950   &44.14   & 0.31 &SNLS\\
171& SN04D3nh    &0.3402    &41.63      &0.17 &SNLS & SN04D3nh   &0.340   &41.51  &0.17 &SNLS\\
172& SN04D4an    &0.6130    &43.08      &0.39 &SNLS & SN04D4an   &0.613   &43.15   &0.18 &SNLS\\
173& SN04D4bq    &0.5500    &42.75      &0.29 &SNLS & SN04D4bq   &0.550   &42.67   &0.17 &SNLS\\
174& SN03D1bp    &0.3460    &41.50      &0.25 &SNLS & SN03D1cm   &0.870  &44.28  &0.34 &SNLS\\
175& SN03D1ew    &0.8680    &43.95      &0.17 &SNLS & SN03D3bh   &0.249  &40.76   &0.17 &SNLS\\
176& SN03D3ba    &0.2912    &40.56      &0.32 &SNLS & SN03D4gl   &0.571   &42.65  &0.18 &SNLS\\
177& SN03D4cn    &0.8180    &44.14      &0.27 &SNLS & SN04D1ag   &0.557   &42.70  &0.17 &SNLS\\
178& SN03D4gf    &0.5810    &42.85      &0.18 &SNLS & SN04D2cf    &0.369   &41.67  &0.17 &SNLS\\
179& SN04D1aj    &0.7210    &43.46      &0.23 &SNLS & SN04D2gp    &0.707  &43.42  &0.21 &SNLS\\
180& SN04D1ak    &0.5260    &42.49      &0.29 &SNLS & SN04D3oe    &0.756   &43.64   &0.17 &SNLS\\
181& SN04D2gc    &0.5210    &42.44      &0.33 &SNLS & SN04D4dm   &0.811   &44.13  &0.31 &SNLS\\
182& SN04D2iu    &0.6910    &43.42      &0.39 &SNLS & SN04D4dw   &0.961  &44.18   & 0.33 &SNLS\\
183& SN04D2ja    &0.7410    &43.58      &0.24 &SNLS& && && \\
184& SN04D3cp    &0.8300    &43.62      &0.17 &SNLS& && &&\\
185& SN04D3dd    &1.0100    &44.70      &0.17 &SNLS& && &&\\
186& SN04D3gt    &0.4510    &41.35      &0.23 &SNLS& && &&\\
187& SN04D3gx    &0.9100    &44.21      &0.18 &SNLS& && &&\\
188& SN04D3is    &0.7100    &43.71      &0.34 &SNLS& && &&\\
189& SN04D3ki    &0.9300    &44.43      &0.19 &SNLS& && &&\\
190& SN04D3ks    &0.7520    &43.36      &0.23 &SNLS& && &&\\
191& SN04D3nc    &0.8170    &43.72      &0.21 &SNLS& && &&\\
192& SN04D4bk    &0.8400    &43.88      &0.17 &SNLS& && &&\\
\hline
\end{tabular}\end{center}

\footnotesize{~~~~~~~~~~~~~~~~~~~Note.--- In the TABLE IV, $^{*}$
denotes the six outliers of the HZSST subset.}

\end{document}